\documentclass[preprint,aps,floats]{revtex4}
\usepackage{epsfig}

\usepackage{graphicx}
\usepackage{dcolumn}
\usepackage{bm}
\def\lsim{\mathrel {\vcenter {\baselineskip 0pt \kern 0pt
    \hbox{$&lt;$} \kern 0pt \hbox{$\sim$} }}}
\def\gsim{\mathrel {\vcenter {\baselineskip 0pt \kern 0pt
    \hbox{$&gt;$} \kern 0pt \hbox{$\sim$} }}}

\newcommand{\U}{{\cal {U}}}

\begin{document}

\title{Supersymmetric Unparticle Effects on \\Higgs Boson Mass and Dark Matter}

\author{N.G. Deshpande$^1$, Xiao-Gang He$^2$ and Jing Jiang$^1$}
\affiliation{ $^1$Institute of
Theoretical Science, University of Oregon, Eugene, OR 97403 \\
$^2$Department of Physics and Center for Theoretical Sciences,
National Taiwan University, Taipei, Taiwan}

\date{\today}

\begin{abstract}
We propose a model that introduces a supersymmetric unparticle
operator in the minimal supersymmetric Standard Model.  We analyze the
lowest dimension operator involving an unparticle. This operator
behaves as a Standard Model gauge singlet and it introduces a new
parameter into the Higgs potential which can provide an alternative
way to relax the upper limit on the lightest Higgs boson mass. This
operator also introduces several unparticle interactions which can
induce a neutral Higgsino to decay into a spinor unparticle. It also
induces violation of scale invariance around the electroweak scale. It
is necessary for the scale of this violation to be larger than the
lightest supersymmetric particle mass to maintain the latter as the
usual weakly interacting massive particle dark matter candidate. An
alternative is to have unparticle state as dark matter candidate.  We
also comment on some collider implications.
\end{abstract}


\maketitle

In a scale invariant theory in four space-time dimensions, the
mass spectrum of fields is zero or continuous. Since in the real
world there are plenty of particles with discrete non-zero masses,
scale invariance, had it existed, must have been
broken at some high energy scale beyond the Standard Model (SM)
scale. At high energy there may be both a scale invariant sector
and a sector which does not obey scale invariance, such
as the SM fields. Recently Georgi proposed an interesting idea to
describe possible scale invariant effects at low energies with an
operator $O_\U$, termed unparticle~\cite{Georgi:2007ek}, which
interacts with the operator $O_{SM}$ composed of SM particles. At
certain scale $\Lambda_{\cal{U}}$ the scale invariant sector with
infrared fixed point induces dimensional transmutation, and below
that scale operators $O_{SI}$ composed of scale invariant fields
matches on to an unparticle operator $O_{\cal{U}}$ with dimension
$d_\U$. The unparticle interaction with SM particles at low energy
has the form
\begin{eqnarray}
\lambda \Lambda_{\cal{U}} ^{4-d_{SM} - d_\U} O_{SM}
O_{\cal{U}}\;.\label{bb}
\end{eqnarray}

The unparticle idea has received a lot of
attention~\cite{Georgi:2007ek, other, Fox:2007sy, Chen:2007qr}.  In
Ref.~\cite{Chen:2007qr} a class of operators involving SM particles
and unparticles are listed. Using these operators one can study
unparticle phenomenology in a systematic way. There have been
extensive studies of unparticle effects at low energies, which can be
used to constrain the unparticle physics scale.

When the scale invariant sector has interactions with the SM sector,
the scale invariance will be broken.  For example an interaction of
the form $\lambda \Lambda^{2-d_\U} H^\dagger H O_\U$ will break scale
invariance, after the Higgs field develops a vacuum expectation value
$v/\sqrt{2}$, since a term of the form $\lambda \Lambda^{2-d_\U}
v^2 O_\U$ will be generated. A Yukawa type of
coupling $ \lambda \Lambda^{-d_\U} \bar Q_L HU_R O_\U$, at one loop
level can generate a term of the form $m^2_\U O^2_\U$ with $m^2_\U$
given by $ m^2_\U \approx ((\lambda \Lambda^{-d_\U})^2/ 16\pi^2)
\Lambda^2_{\mbox{cut}}$. Here $\Lambda_{\mbox{cut}}$ is a cut-off
scale of the effective theory. If the cut-off scale is large, the
breaking of scale invariance can be large. This situation is similar
to the hierarchy problem of Higgs mass. One can eliminate such large
loop correction while maintaining the low energy effect of the
unparticle and stabilize the theory by making the whole theory
supersymmetric.

Supersymmetric theories have many appealing features, they provide a
natural solution to the hierarchy problem. With R-parity they provide
a natural candidate for dark matter. In the minimal supersymmetric
extension of the standard model (MSSM), the lightest Higgs boson mass
is constrained to be less than about 140 GeV.  We anticipate the discovery of
the Higgs boson at the LHC. A mass in the range of 140 GeV is
considered to be consistent with the MSSM.  If a light Higgs boson is
not found, modifications are needed. We find that the supersymmetrized
unparticle effects can relax the Higgs mass limit. This motivates us
to consider unparticle effects in a supersymmetric theory, and we
therefore propose a model for supersymmetrized version of unparticle
interaction.

The model is a minimal extension to MSSM. Besides the usual MSSM
contents with R-parity, we add a complex SM singlet chiral
unparticle operator which has a scalar unparticle $O_\U$ with
dimension $d_\U$ and also a spinoral partner $\tilde O_\U$ with
dimension $d_\U + 1/2$. The associated F-term $F_\U$ has dimension
$d_\U + 1$. Normalizing the supersymmetric unparticle operator to
a dimension one chiral field, we write super-field $Os$ as
\begin{eqnarray}
Os = (O_{s\U} + \theta \tilde O_{s\U} + \theta^2 F_{s\U})~,
\end{eqnarray}
where $(O_{s\U},\;\tilde O_{s\U},\; F_{s\U}) = \Lambda^{1-d_\U}
(O_\U,\; \tilde O_\U,\; F_\U)$. The component super-fields can then be
treated in a similar way to the components of usual chiral fields, in
constructing the supersymmetric Lagrangian.

Since the unparticle does not have gauge interaction, its interactions
with the MSSM particles arise from the super-potential. The lowest
dimension operator involving the unparticle is,
\begin{eqnarray}
\label{eq:lo}
L_O = \lambda H_1 H_2 Os~,
\end{eqnarray}
where $H_{1,2}$ are the two Higgs doublets in the MSSM. The component
fields and the vacuum expectation values (vev's) are, $H_1^T =(h_1^+,
(v_1+ h^0_1 +i a_1)/\sqrt{2}) $, and $H_2^T = ((v_2+ h^0_2 +i
a_2)/\sqrt{2}, h^-_2)$. Depending on the dimension of $O_\U$, the
theory may or may not be renormalizable. Since the theory is an
effective theory, we do not require renormalizability. More
complicated operators can also be introduced. In this letter, we will
concentrate on the effects of this simplest operator.

With this operator, the Higgs potential is given by
\begin{eqnarray}
&&V_{\rm{SUSY}}={1\over 8} (g^2_1+g^2_2) (|H_1|^2-|H_2|^2)^2 + {1\over
2} g_2^2(H_1^\dagger H_2)(H^\dagger_2 H_1) + \lambda^2
|H_1H_2|^2\;,\nonumber\\
&&V_{\rm soft}= \mu^2_1 H^\dagger_1 H_1 + \mu^2_2 H^\dagger_2 H_2 -
\mu^2_{12} (H_1 H_2 + h.c.)\;.
\end{eqnarray}
The term $\lambda^2|H_1H_2|^2$ are obtained by extracting the
F-term from the super-potential $L_O$.

Note that the operator $O_\U$ is similar to the singlet in the
next-MSSM (NMSSM) in gauge properties~\cite{nmssmodel}. However, the
unparticle does not play the role of a massive particle, terms of the
form $O^2_\U$ are not interpreted as mass terms and such terms are
therefore not included in the Higgs potential for Higgs mass
calculations.  Because Eq.~(\ref{eq:lo}) describes an effective
theory, $\lambda$ is not constrained to be small as in the NMSSM.  The
Higgs sector here is very similar to that in $\lambda$MSSM discussed
in~\cite{Barbieri:2006bg}.  Expanding the potential given above, we
obtain the Higgs mass matrices. At the tree level the charged and the
pseudoscalar Higgs boson masses $m^2_{h^{\pm}}$ and $m^2_{A}$ are
\begin{eqnarray}
m^2_A = \mu^2_{12}{v_1^2+v_2^2\over
v_1v_2}\;,\;\;\;\;m^2_{h^{\pm}} = {1\over
4}(g_2^2-2\lambda^2+4{\mu^2_{12}\over v_1v_2})(v_1^2+v_2^2)= m^2_W
(1-2\frac{\lambda^2}{g_2^2}) + m^2_A\;,
\end{eqnarray}
where $g_2$ is the SM $SU(2)_L$ coupling.  The neutral Higgs boson
mass matrix $M^2_h$ is modified by a term proportional to
$\lambda^2$. In the bases $(h^0_1, h^0_2)$, we have
\begin{eqnarray}
M^2_h
= \left(
\begin{array}{cc}
m_Z^2 \cos^2\beta + m_A^2 \sin^2\beta &  -(m_Z^2 + m_A^2 - \lambda^2 v^2)
\sin\beta \cos\beta \\
-(m_Z^2 + m_A^2 - \lambda^2 v^2) \sin\beta \cos\beta & m_Z^2 \sin^2\beta +
m_A^2 \cos^2\beta
\end{array}
\right)~,
\end{eqnarray}
where $v^2 = v_1^2 + v_2^2$ and $\tan\beta = v_2/v_1$.

We note that the pseudoscalar Higgs mass in this model is the same as
that in MSSM. The charged Higgs mass squared is reduced by an amount
of $2\lambda^2 m^2_W/g_2^2$.  In this model, at the one loop level if
small contributions from unparticle loop effects are neglected, the
leading radiative correction to the lightest Higgs boson mass is the
same as that in the MSSM. The main effect is to add~\cite{books}
\begin{equation}
\delta = \frac{1}{\sin^2\beta}\frac{3 G_F m_t^4}{\sqrt{2} \pi^2} \ln
\frac{m_{\tilde t}^2}{m_t^2}
\end{equation}
to the (2,2) component of the neutral Higgs mass matrix, where $G_F$
is the Fermi constant, $m_t$ the top quark mass, and $m_{\tilde t}^2$ 
the top squark mass.

The neutral Higgs mass eigenvalues are now given by
\begin{eqnarray}
m_{h,H}^2 &=& {1\over2} \{m_Z^2 + m_A^2 +\delta \mp [ (m_A^2
+
m_Z^2 -\lambda^2 v^2)^2 \nonumber\\
& -&  4 (m_A^2 -{\lambda^2 v^2\over 2})(m_Z^2 - {\lambda^2
v^2\over 2}) \cos^22\beta -2\delta (m^2_Z-m^2_A)\cos2\beta +
\delta^2 ]^{1\over2} \}\;.
\label{eq:mhH}
\end{eqnarray}

We note that this is the same result for the upper bound on $m_h$ in
the NMSSM.  In particular, for fixed $\lambda$ and $m_A \to \infty$, one
obtains the overall bound~\cite{Drees:1988fc},
\begin{eqnarray}
m^2_h < m^2_Z  \cos^22\beta + \frac{1}{2} \lambda^2 v^2 
\sin^22\beta + \delta \sin^2\beta~.
\label{eq:mssm}
\end{eqnarray}
In our case, we do not restrict $\lambda$ to be necessarily small, or
$m_A$ to be very large.  After this relaxation, we find from
Eq.~(\ref{eq:mhH}) that for a given $m_A$, $m_h$ is maximized when
$\lambda^2 = (m_A^2 + m_Z^2)/v^2$.  The maximum value is
\begin{eqnarray}
m_h^2 &=& m_Z^2 \sin^2\beta + m_A^2 \cos^2\beta \nonumber \\
      &=& - m_Z^2 \cos2\beta + \lambda^2 v^2 \cos^2\beta~,
\label{eq:lambda}
\end{eqnarray}
for $\tan\beta \ge 1$, and
\begin{eqnarray}
m_h^2 &=& m_Z^2 \cos^2\beta + m_A^2 \sin^2\beta \nonumber \\
      &=& m_Z^2 \cos2\beta + \lambda^2 v^2 \sin^2\beta~,
\end{eqnarray}
for $\tan\beta < 1$.  This result is obvious from the $2 \times 2$
Higgs mass matrix.  The highest values of $m_h$ is realized when the
off-diagonal element goes to zero, {\it i.e.} $\lambda^2 = (m_A^2 +
m_Z^2)/v^2$ .  The maximum then equals the smaller of the diagonal
elements.  Since $m_A$ is not restricted in SUSY, the value of $m_h$
can be large as $m_A$ increases.  However, Eq.~(\ref{eq:lambda})
requires that $\lambda$ be correspondingly large.  A very large
$\lambda$ would not be acceptable because radiative effects induced
might be too large.  We assume that $\lambda$ of the order unity would
be acceptable.  The bound in Eq.~(\ref{eq:mssm}) is realized for fixed
value of $\lambda$ and taking the $m_A \to \infty$ limit.  For fixed
value of $m_A$, the limit in Eq.~(\ref{eq:lambda}) is more stringent
than that in Eq.~(\ref{eq:mssm}).  Thus, for a given finite $m_A$, the
Higgs mass limit comes from Eq.~(\ref{eq:lambda}) rather than
Eq.~(\ref{eq:mssm}).

\begin{figure}[htb]
\centering
\includegraphics[width=10cm]{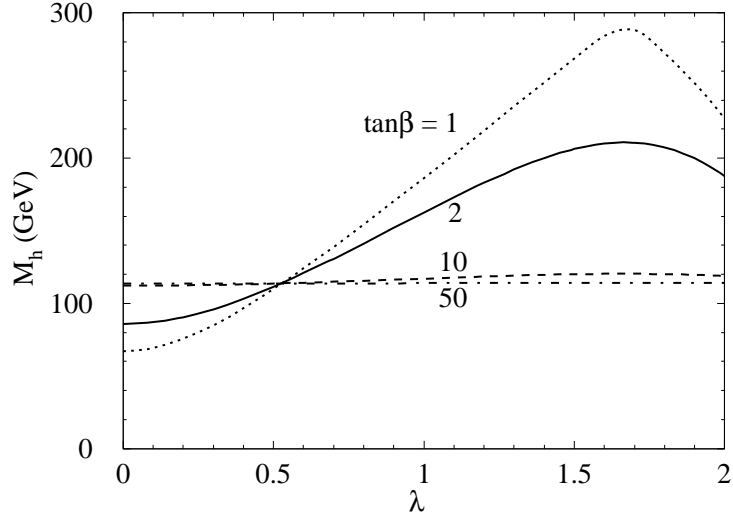}
\caption{The lightest neutral Higgs boson mass as a function of
$\lambda$ for $\tan\beta = 1$, 2, 10 and 50.  The one-loop radiative
correction of top quark is included and $m_A = 400$ GeV and
$m_{\tilde{t}} = 500$ GeV is assumed.} 
\label{fig:hmass}
\end{figure}

In Fig.~\ref{fig:hmass}, we plot the lightest scalar Higgs boson as a
function of $\lambda$ for different values $\tan\beta$.  We take into
account only the radiative corrections due to the top and top-squark
at one-loop level, the mass values will increase with the inclusion of
higher order corrections.  For purpose of illustration, we assume that
$m_A = 400$ GeV and $m_{\tilde{t}} = 500$ GeV.  The dependence of
the top-squark mass is logarithmic.  It is evident that the $\lambda$
dependence is greatest for low value of $\tan\beta$, becoming less
sensitive as $\tan\beta$ increases.  For example, at $\lambda \sim 1.7$
the Higgs mass shift due to $\lambda$ for $\tan\beta = 2$ is about 120
GeV.  The shift for $\tan\beta = 50$ is less than 1 GeV for all values
of $\lambda$.  Note that if we had taken a larger $m_A$, the peak in
Higgs mass would shift to a higher value, and the corresponding value
of $\lambda$ would be higher.  Therefore, if the Higgs boson mass
turns out to be large, the supersymmetric unparticle could be one
possible explanation.

Most of the MSSM interactions will not be affected if the fields do
not couple directly to the Higgs fields. The SM gauge interactions
are not affected. However the new term $L_O$ in the
super-potential brings some interesting consequences. With
R-parity, the lightest supersymmetric particle (LSP) in MSSM is stable
and can be a dark matter candidate. In this model the new operator
introduces several unparticle interactions which may lead
to unstable LSP.  We now discuss this issue in some details.

The new term $L_O$ in the super-potential introduces several new
supersymmetric unparticle interactions with MSSM particles. We
have
\begin{eqnarray}
L_{O_\U} &=& |\mu H_2 +   \lambda \Lambda^{1-d_u}_U H_1 O_u|^2 +
|\mu H_1 + \lambda \Lambda^{1-d_u}_U H_2 O_u|^2 \nonumber \\ 
&& + \lambda\Lambda^{1-d_\U}_\U (\tilde H_1 \tilde H_2 
O_\U + H_1\tilde H_2 \tilde O_\U + \tilde H_1 H_2 \tilde O_\U)\;,
\end{eqnarray}
where the field (operator) with ``tilde'' indicates the
suer-partner field (operator).

A neutral Higgsino can decay into an spinor unparticle due to terms
$H_1\tilde H_2 \tilde O_\U + \tilde H_1 H_2 \tilde O_\U$.  After the
Higgs doublets develop non-zero vev's, a matrix element, $M(\tilde H_i
\to \tilde \U) = \lambda\Lambda^{1-d_\U}_\U v_j\tilde H_i \tilde
O_\U$, is generated. Here $i$ and $j$ take the values 1 and 2 with
$i\neq j$. If the scale invariance of unparticle is unbroken down to
very low energy, the phase space for the unparticle is proportional
$\theta(p^0)\theta(p^2)$, where $p$ is the momentum of the unparticle
and $\theta$ is the step-function. This property allows the Higgsino
of any mass to decay into an unparticle with
\begin{eqnarray}
\Gamma( \tilde H_i \to \tilde \U) = |\lambda \Lambda^{1-d_\U}_\U
v_j|^2 {m_{\tilde H_i}\over 2} A_{d_\U} (m^2_{\tilde H_i})^{d_\U
-2}\theta(m_{\tilde H} )\theta(m^2_{\tilde H})\;,
\end{eqnarray}
where $A_{d_\U} = (16 \pi^{5/2}/(2\pi)^{2d_\U})\Gamma(d_\U +
1/2)/(\Gamma(d_\U -1)\Gamma(2d_\U))$.

It is clear that if the LSP in the MSSM has finite mixing with
Higgsino, it will not be stable in this model. This makes the
conventional explanation of LSP as dark matter candidate untenable.  A
possible solution to maintain all usual weakly interacting massive
particle (WIMP) MSSM dark matter is that the LSP contains no Higgsino
component. This requires fine tuning and is not natural. Here we point
out that the new unparticle interactions introduced in $L_O$ solves
the problem and maintain the LSP to be the usual WIMP dark matter by
itself in a natural way.

The crucial point for this solution is that some of the new
interactions, after the Higgs doublets develop vev's, break scale
invariance explicitly, such as $v_i^2 O_\U^2$ from $|H_i O_\U|^2$
term. One may also introduce a SUSY breaking terms,
$\mu_{\rm{SUSY}}\lambda\Lambda^{1-d_\U} H_1H_2 O_\U$ and
$H^\dagger_iH_iO_\U$ in the theory. These terms induce terms of the
form $v_iv_j O_\U$ which also breaks the scale invariance.  Some
implications for such an operator has been discussed in
Ref.~\cite{Fox:2007sy}. Assuming that the scale for these scale
invariant breaking effects is $\mu^2$, it was suggested in
Ref.~\cite{Fox:2007sy} that the phase space should be changed to be
proportional to $\theta(p^0)\theta(p^2 - \mu^2)$. This implies that
the Higgsino cannot decay into an unparticle if its mass is less than
$\mu$ and would be stable. The relevant scale is proportional to
$v_iv_j$ and is around the electroweak scale. If this is indeed the
case, the LSP in MSSM can still be a good candidate for WIMP dark
matter.  On the other hand, if $\mu$ is less than the mass of the LSP,
the LSP would decay into the lightest supersymmetric unparticle state
and this could be a dark matter candidate.  The properties of dark
matter will be modified, because the lightest supersymmetric
unparticle state has even smaller interaction with ordinary matter.

The new interactions due to $L_O$ can also change Higgs boson
decay properties, in particular increasing the invisible decay rate
for Higgs bosons. For example, the supersymmetric breaking A-term,
$\mu_{\rm{SUSY}}\lambda\Lambda^{1-d_\U} H_1H_2 O_\U$ can induce a term
$\mu_{\rm{SUSY}}\lambda\Lambda^{1-d_\U} (v_i/\sqrt{2} h^0_j) O_\U$
leading to Higgs decay into an unparticle if the Higgs bosom mass
is larger than $\mu$. The decay width is given by
\begin{eqnarray}
\Gamma(h \to \U) = {|\mu_{\rm{SUSY}}\lambda\Lambda^{1-d_\U}|^2\over 2
m_h}\,\left|\frac{v\sin(\alpha+\beta)}{\sqrt{2}}\right|^2 A_{d_\U} (m^2_h)^{d_\U -
2}\theta(m_h)\theta(m^2_h - \mu^2),
\end{eqnarray}
where $\alpha$ is the mixing angle for the neutral Higgs mixing
with $h = \cos\alpha h^0_1 + \sin\alpha h_2^0$ and $H = -\sin
\alpha h^0_1 + \cos\alpha h_2^0$. One can obtain the decay rate
for $H$ by replacing $\sin(\alpha+\beta)$ by $\cos(\alpha+\beta)$
in the above expression.

The term $\lambda^{2}\Lambda^{2-2d_\U}_\U v_ih^0_i O^2_\U/2$
induced from $|H_iO_\U|^2$, will cause Higgs $h^0_i$ to decay into
two unparticles if the Higgs boson mass is larger than $2\mu$.
These decays will contribute to the invisible decay width of Higgs
particle, and affect Higgs search at LHC and ILC. We will present
our detailed analysis elsewhere.

In summary we have proposed an unparticle supersymmetric model by
introducing an supersymmetric unparticle operator to the MSSM. We
introduced the lowest order operator in the super-potential. This
operator helps to relax the upper limit on the lightest Higgs
boson mass. This operator also introduces several unparticle
interactions in the theory leading to interesting phenomenology.
In particular, a neutral Higgsino can decay into an spinor
unparticle. If the LSP in the MSSM has finite mixing with
Higgsino, it will not be stable in this theory and it cannot be a dark
matter candidate. However, the same operator also induces
scale invariant violation at the electroweak scale. If this scale
is larger than the LSP mass, the LSP will again be stable and be a
valid dark matter candidate. The new
interactions can also induce new invisible decay channels and
affect Higgs search at LHC and ILC.

Note Added: While we were finishing our paper, Ref.~\cite{susy}
appeared which also considered some supersymmetric aspects of
unparticle effects.

\vskip 1.0cm \noindent {\bf Acknowledgments}$\,$ This work was
supported in part Grants No. DE-FG02-96ER40949 of the U.S.
Department of Energy, and NSC and NCTS of ROC.


\begin{thebibliography}{99}
\itemsep 0.5mm

\bibitem{Georgi:2007ek}
  H.~Georgi,
  Phys.\ Rev.\ Lett.\  {\bf 98}, 221601 (2007)
  [arXiv:hep-ph/0703260].




\bibitem{other}
  H.~Georgi,
  arXiv:0704.2457 [hep-ph];
  K.~Cheung, W.~Y.~Keung and T.~C.~Yuan,
  arXiv:0704.2588 [hep-ph];
  M.~Luo and G.~Zhu,
  arXiv:0704.3532 [hep-ph];
  C.~H.~Chen and C.~Q.~Geng,
  arXiv:0705.0689 [hep-ph];
  G.~J.~Ding and M.~L.~Yan,
  arXiv:0705.0794 [hep-ph];
  Y.~Liao,
  arXiv:0705.0837 [hep-ph];
  T.~M.~Aliev, A.~S.~Cornell and N.~Gaur,
  arXiv:0705.1326 [hep-ph];
  X.~Q.~Li and Z.~T.~Wei,
  arXiv:0705.1821 [hep-ph];
  C.~D.~Lu, W.~Wang and Y.~M.~Wang,
  arXiv:0705.2909 [hep-ph];
  M.~A.~Stephanov,
  arXiv:0705.3049 [hep-ph];
N.~Greiner,
 arXiv:0705.3518 [hep-ph];
H.~Davoudiasl,
 arXiv:0705.3636 [hep-ph];
D.~Choudhury, D.~K.~Ghosh and Mamta,
arXiv:0705.3637 [hep-ph];
  T.~M.~Aliev, A.~S.~Cornell and N.~Gaur,
  arXiv:0705.4542 [hep-ph];
  P.~Mathews and V.~Ravindran,
  arXiv:0705.4599 [hep-ph];
  S.~Zhou,
  arXiv:0706.0302 [hep-ph];
  G.~J.~Ding and M.~L.~Yan,
  arXiv:0706.0325 [hep-ph];
  C.~H.~Chen and C.~Q.~Geng,
  arXiv:0706.0850 [hep-ph];
  Y.~Liao and J.~Y.~Liu,
  arXiv:0706.1284 [hep-ph];
  M.~Bander, J.~L.~Feng, A.~Rajaraman and Y.~Shirman,
  arXiv:0706.2677 [hep-ph];
  T.~G.~Rizzo,
  arXiv:0706.3025 [hep-ph];
  K.~Cheung, W.~Y.~Keung and T.~C.~Yuan,
  arXiv:0706.3155 [hep-ph];
  H.~Goldberg and P.~Nath,
  arXiv:0706.3898 [hep-ph];
  S.~L.~Chen, X.~G.~He and H.~C.~Tsai,
  arXiv:0707.0187 [hep-ph];
  R.~Zwicky,
  arXiv:0707.0677 [hep-ph];
  T.~Kikuchi and N.~Okada,
  arXiv:0707.0893 [hep-ph];
  R.~Mohanta and A.~K.~Giri,
  arXiv:0707.1234 [hep-ph];
  C.~S.~Huang and X.~H.~Wu,
  arXiv:0707.1268 [hep-ph];
  A.~Lenz,
  arXiv:0707.1535 [hep-ph];
  D.~Choudhury and D.~K.~Ghosh,
  arXiv:0707.2074 [hep-ph];
Xue-Qian Li, Yong Liu, Zheng-Tao Wei,
  arXiv:0707.2285.

\bibitem{Fox:2007sy}
  P.~J.~Fox, A.~Rajaraman and Y.~Shirman,
  arXiv:0705.3092 [hep-ph];

\bibitem{Chen:2007qr}
  S.~L.~Chen and X.~G.~He,
  arXiv:0705.3946 [hep-ph].

\bibitem{nmssmodel}
H.P.~Nilles, M.~Srednicki, and D.~Wyler,
  Phys.\ Lett.\ B {\bf 120}, 346 (1983);
J.M.~Frere, D.R.T.~Jones, and S.~Raby,
  Nucl.\ Phys.\ B {\bf 222}, 11 (1983);
J.P.~Derendinger and C.A.~Savoy,
  {\it ibid}.  {\bf 237}, 307 (1984);
  J.~R.~Ellis, J.~F.~Gunion, H.~E.~Haber, L.~Roszkowski and F.~Zwirner,
  Phys.\ Rev.\ D {\bf 39}, 844 (1989).

\bibitem{Drees:1988fc}
  M.~Drees,
  Int.\ J.\ Mod.\ Phys.\  A {\bf 4}, 3635 (1989).

\bibitem{books}
  M.~Drees, R.~Godbole and P.~Roy,
{\it  Hackensack, USA: World Scientific (2004) 555 p};
  H.~Baer and X.~Tata,
{\it  Cambridge, UK: Univ. Pr. (2006) 537 p}.



\bibitem{Barbieri:2006bg}
  R.~Barbieri, L.~J.~Hall, Y.~Nomura and V.~S.~Rychkov,
  Phys.\ Rev.\  D {\bf 75}, 035007 (2007)
  [arXiv:hep-ph/0607332].



\bibitem{susy}   
  H.~Zhang, C.~S.~Li and Z.~Li,
  arXiv:0707.2132 [hep-ph];
  Y.~Nakayama,
  arXiv:0707.2451 [hep-ph].


\end{thebibliography}
\end{document}